\documentclass[sigconf]{acmart}
\usepackage{booktabs} 

\usepackage{graphicx} 
\usepackage{subfigure} 

\setcopyright{rightsretained}

\acmDOI{}

\acmISBN{}

\acmConference[REVEAL 2018]{Workshop on Offline Evaluation for Recommender Systems}{October 2018}{Vancouver, Canada}
\acmYear{2018}
\copyrightyear{2018}

\acmArticle{4}
\acmPrice{15.00}

\begin{document}
\title{Offline Comparison of Ranking Functions using Randomized Data}

\author{Aman Agarwal, Xuanhui Wang, Cheng Li, Michael Bendersky, Marc Najork}
\affiliation{%
  \institution{Google}
  \streetaddress{}
  \city{Mountain View} 
  \state{CA} 
  \postcode{94043}
}
\email{{agaman, xuanhui, chgli, bemike, najork}@google.com}

\renewcommand{\shortauthors}{A. Agarwal et al.}

\begin{abstract}
Ranking functions return ranked lists of items, and users often interact with these items. How to evaluate ranking functions using historical interaction logs, also known as off-policy evaluation, is an important but challenging problem. The commonly used Inverse Propensity Scores (IPS) approaches work better for the single item case, but suffer from extremely low data efficiency for the ranked list case. In this paper, we study how to improve the data efficiency of IPS approaches in the offline comparison setting. We propose two approaches \emph{Trunc-match} and \emph{Rand-interleaving} for offline comparison using uniformly randomized data. We show that these methods can improve the data efficiency and also the comparison sensitivity based on one of the largest email search engines.
\end{abstract}

%
%



\maketitle

\section{Introduction}

Recommender systems or search engines usually return a list of items, ordered by underlying ranking functions, to end users. While human labeled data can be used to evaluate ranking functions in an offline setting, such data is not always available at large scale and also expensive to collect and maintain. On the other hand, A/B experiments based on user interaction data from production traffic are commonly used for evaluation in an online setting. However, they are costly to set up and the comparison results can be difficult to reuse~\cite{Li:WSDM2015,Swaminathan:NIPS2017}. How to compare ranking functions in an offline setting based on historical user interaction logs is an active research area.

Causal inference or counterfactural framework is commonly used to leverage historical logs to evaluate new ranking functions in an offline setting (e.g., \cite{li2011unbiased, bottou2013counterfactual}). However, most of this work focuses on the single item case, but not the ranked list one. Recently, there is an increasing research interest~\cite{Li:WSDM2015,Swaminathan:NIPS2017} in offline evaluation of ranked lists (also called slates). Among them, Inverse Propensity Scores (IPS) approaches are commonly used. The main challenge in IPS approaches is that there are $n!$ possible ranked lists for a set of $n$ items. The chance of matching a new ranked list against the evaluation logs becomes tiny when $n$ becomes large. This incurs very low data efficiency. To increase the data efficiency, Li et al.~\cite{Li:WSDM2015} and Wang et al.~\cite{Wang:SIGIR2016} proposed a partial matching method that only matches the top $k$ items for two ranked lists. How to further improve the data efficiency is the topic of this paper.

In this paper, we focus on improving data efficiency in IPS approaches. One of the main characteristics of existing methods using exact (partial) matching is that it can in general give point estimation of the metrics of interest such as CTR. However, in this paper we focus on the \emph{comparison} scenario where we are only interested in whether a ranking function is better or worse than another. Such a deviation gives us more flexibility to design offline evaluation with higher data efficiency. Interestingly, the observation that pairwise comparison is more data-efficient than pointwise evaluation is also made in \cite{Schnabel:2016}, which addresses the different but complementary problem of designing the optimal sampling distribution for the data collection step given the ranking functions to be compared. 

Specifically, we propose the following two approaches based on uniformly randomized data. The first approach \emph{Trunc-match} uses a truncated version of randomized data to improve the matching ratio. The second approach \emph{Rand-interleaving} is based on the interleaving methodology to compare two ranking functions at a time. Interleaving was developed as an alternative for online A/B experiments for ranking problems and was shown to be more sensitive~\cite{Radlinski:CIKM2008} in comparison. To the best of our knowledge, this paper is the first to use it for offline evaluation.

In the following, we first describe our proposed approaches and then report our evaluation results based on a large-scale commercial email search engine.
\section{Methods}

Using randomization during data collection has been proposed as a reliable way for offline evaluation~\cite{li2011unbiased, bottou2013counterfactual}. In order to do this, an experiment needs to be run on a small fraction of user traffic, from which randomized data can be collected. During the period of the experiment, when a ranked list is returned by the production ranker, the top $n$ results are randomly shuffled before being presented to users. The presented results and user interactions such as clicks are collected to form the randomized data set. Now given a new ranker $A$, we are able to evaluate its performance offline using the randomized data collected by the above procedure. In the following, we use $n$ to denote the number of items in randomized data and $k$ $(k \leq n)$ to denote the top $k$ of the $n$ items. We also assume that our data is uniformly randomized.

\subsection{Direct-match}
The conventional procedure, which we call \textbf{Direct-match}, works as follows based on uniformly randomized data. First, rank all $n$ items based on ranker $A$. Second, compare the top $k$ results from ranker $A$ with the top $k$ results recorded in the randomized data. If both top $k$ results are matching exactly, keep the ranked lists.

At the end, we obtain a subset of the randomized data based on the Direct-match procedure. Evaluation metrics like mean reciprocal rank (MRR) can be computed for ranker $A$ based on this subset. Such a method is provably unbiased~\cite{Li:WSDM2015,Wang:SIGIR2016}.

\subsection{Trunc-match}
The \textbf{Direct-match} method provides an unbiased evaluation of any ranker in an offline manner. However, its data efficiency is low, as a large fraction of randomized data has to be discarded when their recorded lists mismatch the ranked lists of a ranker to be evaluated. For example, suppose that we have 6 results in a ranked list, then the chance that we can match the top $k=3$ items is $\frac{1}{6 \cdot 5 \cdot 4} = \frac{1}{120}$. In order to alleviate this problem and improve data efficiency, we propose \textbf{Trunc-match}, which is a minor modification of the Direct-match method.

Specifically, given $k$, we first truncate the recorded lists in the randomized data to keep the top $k$ results for each list. It can be proved that such a truncated data set is still uniformly randomized. Now given a ranker $A$, instead of evaluating it over the original data, we only have it rank the top $k$ documents in the truncated set. The remaining procedure is the same as Direct-match -- we collect the matched lists, and calculate evaluation metrics.

In this way, we will have $\frac{1}{3 \cdot 2 \cdot 1} = \frac{1}{6}$ matching probability when $k=3$, which greatly improves data utilization compared with Direct-match. The downside is that such a method can not give the point estimation of the metrics of interest. This is not a concern when our main focus is on \emph{comparing} different rankers.

\subsection{Rand-interleaving}
In the comparison case we are more interested in knowing the relative performance of two rankers, instead of the absolute performance of a single ranker. This leads us to design an interleaving method for offline comparison.

Given $k$ items, ranker $A$ and ranker $B$ give two ranked lists of the $k$ items. We use the balanced interleaving approach in~\cite{Radlinski:CIKM2008} to obtain a single ranked list and then match it against the recorded one in the randomized data set. We choose to use the Trunc-match method to maintain the high data efficiency. After matching, we use the same logic as the balanced interleaving to attribute the recorded clicks to the two rankers and then compare them based on the resultant attributed clicks. We name our method \textbf{Rand-interleaving} to distinguish it from the conventional online interleaving methods. This interleaving method is less sensitive to data variance, since the two rankers can be directly compared for each ranked list. 


\section{Experiments}

\subsection{Data Sets}
We evaluate our proposed methods based on one of the largest commercial email search engines. In this service, there is at most one single click for each query. This is because the service uses an overlay to show the results as users type and the overlay disappears when a click on the overlay happens. The data set we used in this paper is the randomized data collected in a two-week period of December 2017. Given queries, rankings were presented uniformly at random. Overall, the randomized data has 1,034,343 queries. Each query has around 5 results.

\subsection{Evaluation Metrics}
We compare our offline evaluation methods Trunc-match and Rand-interleaving against the baseline Direct-match using the randomized data. For comparison, we select two internally designed ranking functions whose qualitative relative performance is known. The task then is to assess whether our proposed evaluation methods can correctly and with statistical significance indicate which ranking function is better using the existing randomized data.

We consider the Mean Reciprocal Rank (MRR) metric at a range of top positions, denoted as $MRR@k$, for Direct-match and Trunc-match, and the number of clicks for Rand-interleaving. In order to obtain metric estimates with error bars, we resample multiple $50\%$ slices of the randomized data set, and report the mean and standard error of the estimates from each evaluation method.

\subsection{Experimental Results}
We first analyze how well the methods match the rankings for the sampled queries. High retention rate of the queries indicates the reliability of the evaluation methods. In Table $\ref{tab:queries}$, we see that both Trunc-match and Rand-interleaving retain substantially more queries than Direct-match. Moreover, as expected, Trunc-match retains about $\frac{1}{k!}$ queries of the $50\%$ samples. Interestingly, Rand-interleaving succeeds in retaining slightly more queries than Trunc-match. While we omit a formal proof, intuitively, this is because the probability of matching each position in a random ranking is higher with two models (interleaved by taking minimum over ranks from each model) than with each model individually. Also, as expected, we observe that the number of queries retained decreases with increasing $k$ for all methods due to the growing number of positions to be matched.

\begin{table}[h]
    \begin{tabular}{rrrrr}
        \toprule
        $k$   & \# Direct-match & \# Trunc-match & \# Rand-interleaving \\
        \midrule
        1   & 134,568 & 516,514 & 515,941 &   \\
        2   & 35,858 & 248,924  & 249,190 &   \\
        3   & 8,886 & 80,786 & 80,879 &    \\
        4  & 2,290 & 19,696 & 19,705  &  \\
        \bottomrule
    \end{tabular}
    \caption{Average number of queries retained using each method from $50\%$ random slices of the 1,034,343 queries in the randomized data set.}
    \label{tab:queries}
\end{table}

\begin{figure}
    \centering
    \includegraphics*[width=0.9\linewidth]{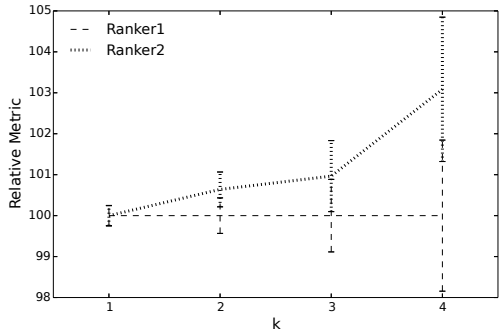}
    \vspace*{-0.3cm}
    \caption{Direct-match}
    \label{fig:match}
\end{figure}

\begin{figure}
    \centering
    \includegraphics*[width=0.9\linewidth]{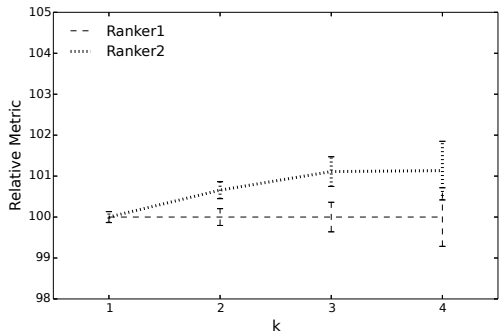}
    \vspace*{-0.3cm}
    \caption{Trunc-match}
    \label{fig:truncmatch}
\end{figure}

\begin{figure}
    \centering
    \includegraphics*[width=0.9\linewidth]{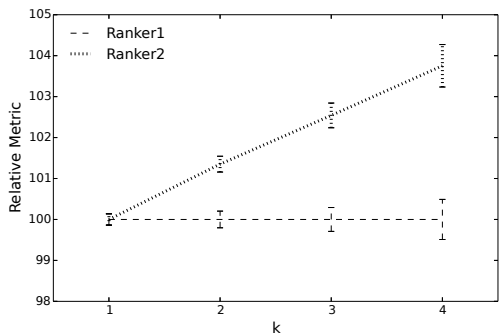}
    \vspace*{-0.3cm}
    \caption{Rand-interleaving}
    \label{fig:randinter}
\end{figure}

Now we compare the evaluation methods by using them to compare two rankers. Since we cannot disclose the absolute value of metrics (i.e., MRR@k or Clicks), we report the relative performance of Ranker 2 (better ranker) against Ranker 1 (worse ranker), by always setting the value of Ranker 1 to 100. The results for each method is shown in Figure~\ref{fig:match}, \ref{fig:truncmatch}, and \ref{fig:randinter} by varying $k$.

There are some commonalities among the three methods. First, the error bars tend to grow larger as the value of $k$ increases, which is as expected. Second, all methods show that Ranker 2 outperforms Ranker 1, verifying the correctness of these evaluation methods.

Comparing among the three methods, the error bars of Direct-match overlap even at small values of $k$, empirically confirming its inefficiency of data utilization. Trunc-match performs better than Direct-match, while Rand-interleaving separates Ranker 1 and 2 best both in terms of mean and standard error. This shows that the big advantage of interleaving is not just from matching slightly more queries, but fundamentally from being more sensitive by considering both rankers in each query.

\section{Conclusion}\label{sec:conclu}
In this paper, we proposed two methods, Trunc-match and Rand-interleaving, to improve data efficiency and offline comparison sensitivity using randomized data. Our experimental results on a large-scale commercial email search engine demonstrated the effectiveness of our proposed methods.

Our work can be extended in the following directions. (1) For simplicity, we worked with uniformly randomized data in this paper. It is interesting to extend our approaches to general non-uniformly randomized data or from multiple loggers~\cite{Agarwal/etal/17a, Carterette:2018}. (2) There is a wealth of existing literature on various interleaving methodologies for online comparison of ranking functions (see \cite{hofmann2016online} for a comprehensive survey). Our paper introduced the basic interleaving method in an offline setting. Thus, a natural direction is to study those more sophisticated interleaving methodologies in offline evaluation of ranking functions. (3) Unbiased learning-to-rank~\cite{Joachims:WSDM17,Wang:SIGIR2016,Wang+al:2018,Ai/etal/18a} employs IPS to correct click bias and it would be interesting to see how the proposed evaluation methods can be effectively applied to this set of problems.

\bibliographystyle{ACM-Reference-Format}
\bibliography{bibliography}

\end{document}